\documentclass[%
 reprint,
 amsmath,amssymb,
 aps,
]{revtex4-1}

\usepackage{graphicx}
\usepackage{dcolumn}
\usepackage{bm}

\begin{document}
\preprint{APS/123-QED}
\title{Polarization conversion loss in birefringent crystalline resonators}
\author{Ivan S. Grudinin}
 \email{grudinin@jpl.nasa.gov}
\author{Guoping Lin}%
\author{Nan Yu}%
\affiliation{%
 Jet Propulsion Laboratory, California Institute of Technology,\\
 4800 Oak Grove dr., Pasadena, CA 91109, USA 
}%
\begin{abstract}
Whispering gallery modes in birefringent crystalline resonators are investigated. We experimentally investigate the XY--cut resonators made with LiNbO$_3$, LiTaO$_3$ and BBO and observe strong influence of the resonator's shape and birefringence on the quality factor of the extraordinary polarized modes. We show that extraordinary modes can have lower Q and even be suppressed due to polarization conversion loss. The ordinary ray modes retain the high Q due to inhibited reflection phenomenon.
\end{abstract}

\maketitle


\noindent Crystalline whispering gallery mode (WGM) resonators are known for compact size and high optical Q factors. They have been used to enhance a broad range of quantum, optomechanical and nonlinear optical processes \cite{ilchrev}.
In nearly all published work, the WGM resonator is fabricated with its axis aligned with the crystalline optical axis. Such Z--cut resonators are experimentally known to support two families of whispering gallery modes having polarization along the resonator axis (here called TE modes) and orthogonal to it (TM modes). The nonlinear frequency conversion efficiency in a WGM resonator depends on the component of the nonlinear susceptibility tensor \cite{boyd}. This component is determined by the WGM polarizations and crystalline orientation. For this reason the use of X-- and Y--cut resonators \cite{quartz}, or angle--cut resonators \cite{guoping}, may be beneficial. The X-- and Y--cut resonators made of uniaxial crystals are optically equivalent in the linear optics regime. In the case of X--cut resonators the TE modes experience constant ordinary refractive index $n_o$, and the TM modes experience varying extraordinary index $n_e$.

In this paper we present the analysis of WGMs based on the laws of reflection from the boundary separating a birefringent crystal and air. We predict general properties of WGM resonators made with the birefringent crystals and carry out experiments to support our analysis. We show that there exists a loss mechanism that limits the Q factor of the extraordinary polarized modes up to complete suppression.

\begin{figure}[htb]
\centerline{\includegraphics[width=6.0cm]{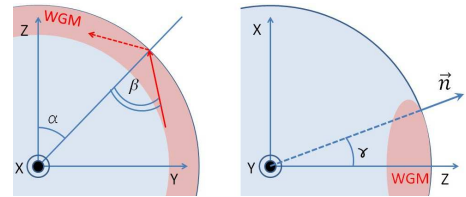}}
\caption{\label{fig:angles} Schematics of an X--cut resonator and reflection angles. (Color online)}
\end{figure}

A number of methods can be used to analyze the WGMs in isotropic dielectrics 
\cite{spheroid,raychaos, fdtdwgm}. However, none of these generally applies to crystalline resonators due to anisotropy. Reflection of either ordinary or extraordinary beams from the crystal-air interface is described by five coefficients \cite{fresnel} as it generally produces one transmitted beam ($T_o, T_e$) and two reflected beams --- ordinary ($R_{oo},R_{eo}$) and extraordinary ($R_{oe},R_{ee}$). The Fresnel coefficients of this process depend on the incident Pointing vector direction, the electric vector polarization, the orientation of the crystalline optical axis, and the direction of the interface normal. For example, in a 1 mm dielectric sphere with the refractive index n=2.12 the half-maximum level of the fundamental WGM field near vacuum wavelength of 1.56 microns will span the meridional angle of 1.4$^\circ$ (e.g. Fig. \ref{fig:angles}). The portions of the WGM that locally experience different surface normal directions will experience different Fresnel reflection coefficients .
Thus, direct usage of a ray tracing method is not justified. While a general method is missing, we can still rely on the anisotropic Fresnel coefficients \cite{fresnel} computed numerically to analyze some general properties of crystalline resonators.

\begin{figure}[htb]
\centerline{\includegraphics[width=8.5cm]{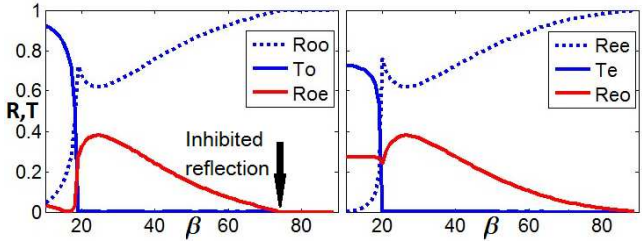}}
\caption{\label{fig:inhibited} Reflection and transmission coefficients for $\alpha=0^{\circ}, \gamma=20^{\circ}$ (chosen to exaggerate the non-isotropic behavior) as a function of incidence angle $\beta$ in LN ($n_e$=2.12, $n_o$=2.21). (Color online)}
\end{figure}

For simplicity, we consider reflections from the boundary of a spherical resonator made with a negative crystal as shown in Fig. \ref{fig:angles}. The boundary is approximated as a plane with normal vector $\vec{n}$ at each reflection point along the curved surface. Here $\alpha$ is the angle between the optical axis (Z) and the radius vector of the WGM reflection point, $\beta$ is the WGM incidence angle in geometric optic approximation and $\gamma$ is the angle between the interface normal and the Y-Z plane.
For the reflection point exactly in the YZ plane where $\gamma=0$, the computation yields the isotropic Fresnel equation behavior for the reflection and transmission coefficients based on $n_o$ for the ordinary ray and local value of $n_e(\alpha,\beta)$ for the extraordinary ray. In this special case both rays experience total internal reflection and Brewster angle exists for the extraordinary polarized ray. The reflection coefficients from extraordinary to ordinary ($R_{eo}$) and ordinary to extraordinary ($R_{oe}$) are zero.
For $\gamma\not =0$, the reflection laws are more complicated. We illustrate this in Fig. \ref{fig:inhibited} by calculating the reflection and transmission coefficients for an X cut LiNbO$_3$ (LN) disk.
One can see that both ordinary and extraordinary beams are created from a single incident beam.  However, there exists a characteristic incidence angle of inhibited reflection $\beta_i(\alpha, \gamma)$ such that for $\beta>\beta_i$, $R_{oe}=0$. Numerically we obtain $\beta_i(0^{\circ},1^{\circ})\simeq 74.5^{\circ}$, which can also be found \cite{inhibited} as $\beta_i=arcsin(n_e/n_o)\simeq 74.5^{\circ}$ for $\alpha=0$. Thus, ordinary polarized TE modes in X cut resonators can still have high Q factors due to the inhibited reflection phenomenon, i.e. inhibited polarization conversion.

\begin{figure}[htb]
\centerline{\includegraphics[width=8.5cm]{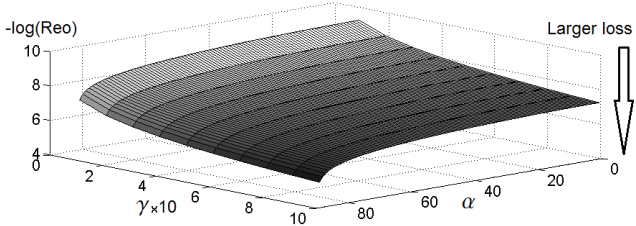}}
\caption{\label{fig:loss} $-log(R_{eo})$ for a 4 mm LN resonator near 1550 nm wavelength ($l=17000,\beta=89.9895^\circ$). $R_{eo}$ grows with $\gamma$ and peaks when $\alpha$ approaches $\pi/2+N\pi, (N=0,1,2...)$. The higher values on the vertical axis correspond to smaller value of $R_{eo}$.}
\end{figure}

For $\gamma\not= 0$, $R_{eo}$ as a function of ($\alpha,\gamma, n_o, n_e$) is always positive under total internal reflection conditions of the WGMs (Fig. \ref{fig:inhibited}). Thus, for the extraordinary polarized TM modes in X cut resonators, there is a polarization conversion loss resulting in reduced Q factor. However, the precise relation between this loss and the WGM quality factor is beyond the scope of this investigation. Qualitatively, conversion loss $R_{eo}$ grows with $\gamma$, hence one may expect that the higher order extraordinary polarized WGMs will be suppressed. Even the fundamental mode can be suppressed if it experiences reflection from the surface with locally significant $\gamma$. We compute $R_{eo}$ for an X cut LN resonator geometry for a range of $\alpha$ and $\gamma$ angles as shown in Fig. \ref{fig:loss}. 
Here the incidence angle is chosen as $\beta=\pi/2-\pi/l$, where $l$ is the azimuthal WGM order. To investigate how birefringence influences $R_{eo}$, we calculate the ratios of $R_{eo}$ of LN and beta-barium borate (BBO, $n_e$=1.54, $n_o$=1.65) to that of lithium tantalate (LT), which has near zero birefringence. $R_{eo}(LN)/R_{eo}(LT)$ varies between 6 and 7, while $R_{eo}(BBO)/R_{eo}(LT)$ varies between 11 and 12 for the same range of $\alpha$ and $\gamma$ angles as in Fig. \ref{fig:loss}.
Based on the behavior of $R_{eo}$ we can predict that the TM modes in X cut resonators and extraordinary polarized modes in generalized angle cut resonators will be inhibited. The degree of Q factor degradation will depend on the reflection angles $\gamma$ as experienced locally by a WG mode and on the birefringence. The TE modes in Z cut resonators should also experience Q factor degradation, albeit to a much lesser degree.

\begin{figure}[htb]
\centerline{\includegraphics[width=7.0cm]{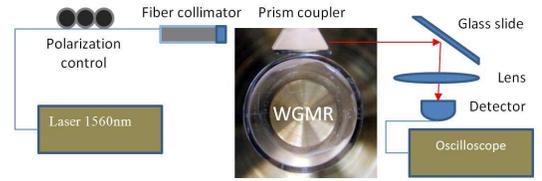}}
\caption{\label{fig:setup} Schematics of the experimental setup. A 4 mm in diameter WGM resonator and a prism coupler are shown on the photo inset. (Color online) }
\end{figure}

\begin{figure}[htb]
\centerline{\includegraphics[width=7cm]{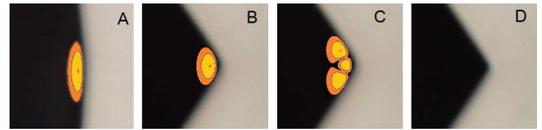}}
\caption{\label{fig:fem} FEM modeling of the TE modes of the dielectric resonators A) First order mode in the LN resonator, 170$^\circ$ edge angle. B,C) First and third order modes in the LN resonator, 110$^\circ$ edge angle. D) Profile of a LT resonator. Each window is $100\times 100$ $\mu m$. Yellow area boundary corresponds to half of the field maximum, orange to the 1/e level. (Color online)}
\end{figure}

\begin{figure}[htb]
\centerline{\includegraphics[width=8.0cm]{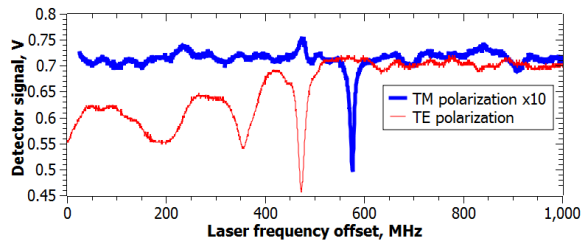}}
\caption{\label{fig:TETM}A portion of the spectrum of the 4 mm multimode LN resonator for TE and TM polarizations. The TM signal is multiplied tenfold due to smaller reflection coefficient of the glass slide. While TM mode is critically coupled, the TE modes are strongly overcoupled. (Color online)}
\end{figure}

To experimentally investigate the polarization conversion loss and its impact on WGM Q factor we used X cut LN and LT wafers to fabricate two pairs of 4 mm diameter resonators. One pair of the resonators has 170$^\circ$ edge angle (multimode resonators) and the other 110$^\circ$ (single mode resonators). The modes were excited with a diamond prism coupler, using a fiber collimator based on a GRIN lens. The laser frequency was scanned with a piezo drive and could also be temperature tuned around 1560-1561 nm. Schematics of the setup are shown in Fig. \ref{fig:setup}.
A glass slide was used as a polarization analyzer. The TE modes could be excited with arbitrary Z axis orientation with respect to the prism surface, as ordinary refractive index is a constant. To couple to TM modes, the collimator launch angle was adjusted for the selected disk orientation. We used a publicly available FEM solver \cite{freefem} to obtain the WGM patterns in our resonators as shown in Fig. \ref{fig:fem}, treating the crystal as an isotropic dielectric. 
In the multimode X cut LN resonator, where the resonator edge is relatively blunt, we observed a family of high Q TE modes with material limited intrinsic quality factor $Q=2.2\times 10^8$. Upon changing the polarization to excite the TM modes we found a much smaller number of modes with intrinsic Q factors up to $Q=3\times 10^7$ as shown in Fig. \ref{fig:TETM}. Maximum coupling for this resonator was 35\% for both polarizations. However, due to the lower Q factor of TM modes, the gap between the prism and the disk had to be reduced to observe these modes in critically coupled regime. The reduced gap resulted in overcoupling condition for the TE modes in Fig. \ref{fig:TETM}.
Similarly, we fabricated an XY--cut (only Z axis is known to be in WGM plane) BBO resonator with 1.73 mm diameter and about 150 micrometers radius of curvature at the resonator edge. We observed that the  Q factor for the TM modes near 633 nm wavelength was suppressed by a factor of 4 ($Q_{TE}=3.6\times 10^7$, $Q_{TM}=1\times 10^7$). 

In a single mode LN resonator with relatively sharp, 110$^\circ$ edge we observed one high--Q TE mode with a coupling of around 20\% and two other modes with much weaker coupling. However, we observed no modes in TM polarization at the level of detector noise with all possible coupling configurations. We also fabricated a 4 mm diameter LN X cut resonator with edge angle of nearly 176$^\circ$ and found the TM modes with Q factor of up to $4\times 10^7$, indicating reduced conversion loss compared to the resonator with 170$^\circ$ edge. 

The multimode LT resonators were found to have clearly defined TE and TM mode families with similar coupling and material limited intrinsic Q factors as high as 100 million. In a single mode LT resonator we found the intrinsic Q factors of around 50 million for both TE and TM modes. 

These experimental observations strongly suggest that the degree of Q factor degradation for the extraordinary polarized modes depends on the resonator shape and birefringence. In X--cut resonators made of negative crystals the Q factor of TM modes degrades with the decrease in resonator's edge radius and increase in birefringence. 

In general, the inhibited reflection phenomenon \cite{inhibited} explains the high Q factor of the ordinarily polarized modes in the resonators made with negative crystals.
The geometry-dependent polarization conversion loss is present for the extraordinarily polarized WG modes in negative crystals. This loss explains the absence of extraordinary modes in the extreme case of angle cut BBO resonators \cite{guoping}. 

In positive crystals ($n_e>n_o$) the inhibited reflection takes place for the extraordinary ray and the unaffected modes would be TM polarized in an X--cut resonator. Along these lines, in positive crystals the Q factor of the ordinary polarized TE modes in an X--cut disk may be degraded depending on resonator geometry. In biaxial crystals there are no ordinary rays. Thus one can expect geometry dependent polarization loss for all kinds of WG modes in biaxial crystalline resonators.

In conclusion, we show analytically and experimentally that there exists the polarization conversion loss in WGM resonators made with anisotropic crystals. In negative crystals such as LN, LT and BBO, the birefringence induced polarization conversion loss suppresses the TM modes in XY-- and angle--cut resonators. Our work paves the way towards better understanding of polarization conversion loss and helps design birefringent crystalline non Z--cut resonators supporting both TE and TM modes having high optical Q factors. Further quantitative study and the development of the FEM, ray tracing or other methods generalized for the case of geometry--dependent WGMs in anisotropic dielectric resonators will enable better resonator engineering and its applications.

This work was carried out at the Jet Propulsion Laboratory, California Institute of Technology under a contract with the National Aeronautics and Space Administration, with support from NASA Center Innovation Fund and JPL Research and Technology Development Program.

\end{document}